\documentclass{camera}
\usepackage{graphicx}  % uncomment this if your want to insert figures
\input psfig.sty
\input epsf.sty

\def \asca {\hbox{\it ASCA }}

\def \msun {\hbox{$M_{\sun}$}}

\def \xmm {{\it XMM-Newton} }
\def \sax {{\it Beppo-SAX} }

\def\cha{{\em Chandra  }}
\def\xmm{{\em XMM-Newton }}
\def\msun{M$_\odot$}

\def \h5-1 { $h_{50}^{-1}$}

\begin{document}

%%%%%%%%%%%%%%%%%%%%%%%%%%%%%%%%%%%%%%%%%%%%%%%%%%%%%%%%
% The title, only the first letter capitalized; if you want to split it in
% two or more lines, put a \\ macro at each line break
% example: 
%   \title{Title: first line\\ second line}
%
\title{New results on clusters of galaxies observed with \xmm}

%%%%%%%%%%%%%%%%%%%%%%%%%%%%%%%%%%%%%%%%%%%%%%%%%%%%%%%%
% The author(s), separated by commas; do not put a
% comma before the last author, use instead the \and
% macro which produces a normal ``and'' in the
% caps/small caps context
%
\author{Doris M. Neumann \and S. Majerowicz}

%%%%%%%%%%%%%%%%%%%%%%%%%%%%%%%%%%%%%%%%%%%%%%%%%%%%%%%%
%
\organization{CEA/Saclay DSM/DAPNIA/Service d'Astrophysique, France}

\maketitle

\begin{abstract}

Clusters of galaxies contain a hot gas, which emits in X-rays. X-ray telescopes such as \xmm allow to study this plasma to obtain information on physical quantities of these objects. We present here some results on the total mass density distribution of clusters obtained with \xmm based on the hydrostatic approach. These results can be compared to models based on cold dark matter. Generally good agreement is found between observations and models. Furthermore we present a study on physical properties of a distant merging cluster of galaxies, which demonstrates the potential of \xmm studies on this class of objects.  

\end{abstract}

%%%%%%%%%%%%%%%%%%%%%%%%%%%%%%%%%%%%%%%%%%%%%%%%%%%%%%%%
% Write the text starting from here and using the usual
% LaTeX commands.
%

\section{Introduction}

Clusters of galaxies comprise total masses which often exceed several $10^{14}$ up to a few $10^{15}$  \msun and typical sizes of a few Mpc. Clusters belong therefore to the largest gravitationally bound objects which attained so far equilibrium state (see for example \cite{bahcall}). 

In the context of hierarchical structure formation, in which small objects collapse before bigger ones, clusters of galaxies belong to the youngest structures in the universe we can observe today.  We can witness cluster formation even in the present day universe: galaxy clusters exhibit to a large fraction substructure (see for example \cite{schuecker}), which can be identified as the merging with other groups of galaxies or clusters.

Because of their relatively young age internal physical processes in clusters did not have time to erase information on the initial over-density and collapse from which they originate. This allows us to use clusters of galaxies as probes to study structure formation, and at the same time for the determination of cosmological parameters (see for example \cite{rosati} for a review). 

Clusters of galaxies are dominated by dark matter, which represents about 70-80\% of their total mass. Roughly 20\% of the total mass can be found in the hot intra-cluster medium (hereafter ICM). This is a thin plasma (the central gas number density varies between $10^{-2}-10^{-4}$cm$^{-3}$) with a temperature between typically $10^7-10^8$~K, which can be observed in X-rays (see \cite{sarazin} for a review). The rest of the mass (about 5\%) can be found in galaxies. Beside this mass constituents there exists also a relativistic non-thermal component in many clusters, however its mass contribution is negligible.  

The X-ray observatory \xmm, which is sensitive in the energy 
range 0.1-15~keV \cite{jansen} allows us to study the ICM in
unprecedented detail. Measuring the morphology and temperature distribution 
enables us to determine the total mass of the cluster via the hydrostatic
approach by treating the gas as an atmosphere. Furthermore, investigations of
the ICM allow us to determine the dynamical state of clusters. For
example the infall of substructures onto clusters can create shock waves due
to velocities which exceed the sound speed of the ICM. These shock
waves heat the ICM locally and can therefore be observed by strong variations 
in the temperature distribution observable with modern X-ray telescopes such as \xmm or \cha. 

The paper is outlined as followed: after the introduction we briefly introduce \xmm. This is followed by a section presenting recent results on mass determination of clusters of galaxies and its impact on current models of dark matter. After this we present briefly the work on \xmm data of a distant merging cluster of galaxies.

\section{Introduction to \xmm}

\xmm is an X-ray satellite which was launched into orbit in December 1999 by
an Ariane V rocket.

\xmm has three X-ray telescopes\footnote{These are Wolter type I telescopes,
which allow reflection of X-ray photons based on grazing incident angles.}
(for an overview of \xmm see \cite{jansen}). Two of the three telescopes have X-ray gratings in  the optical path, which act simultaneously as beam splitters \footnote{For figures of the optical path of \xmm as well as its design see also : {\small \tt http://xmm.vilspa.esa.es/external/xmm$_-$user$_-$support/documentation/build}}. A photon passing one of the beam splitters falls either onto the RGS (Reflection Grating Spectrometer) camera \cite{denherder}, which provides high spectral resolution capabilities, or the photon is detected by the MOS-camera \cite{turner}, which provides at the same time 
spatial and spectral resolution. This resolution, however, is of much lower quality than the one obtained by the RGS camera. The last X-ray telescope has only one detector in its focal plane. This is the pn-camera \cite{strueder}, which 
consists of back-illuminated CCD's and which offers, as the MOS-cameras spatial and spectral resolution at the same time. 

The two MOS-cameras together with the pn-camera form the so-called European 
Photon Imaging Camera (EPIC) \cite{turner}. The two RGS cameras with high spectral resolution are described in \cite{denherder}.

Apart from the X-ray telescopes \xmm hosts also a small optical telescope
aboard, the optical monitor (see \cite{mason} for an introduction).

\section{The mass density distribution and the origin of dark matter}

Clusters are dark matter dominated and offer therefore a unique possibility to study this kind of material. The currently favoured idea is that dark matter is ``cold'' (the so-called cold dark matter -- CDM). Cold refers to the fact that the particles are non-relativistic. Numerical simulations on CDM  halos predict a unique density profile. There are currently two density profiles proposed. The first \cite{NFW}, the so-called NFW-profiles has the form:

\begin{equation}
\rho_{DM} \propto \frac{r_s}{r(1+r/r_s)^2}
\end{equation}

$\rho_{DM}$ is the dark matter density, and $r_s$ is the so-called scale radius.
The second profile proposed by \cite{moore} has the form:

\begin{equation}
\rho_{DM} \propto \left(\frac{r_s}{r(1+r/r_s)}\right)^{3/2}
\end{equation}

In both profiles the density profile does not show a core region with constant density but instead $\rho_{DM}$ goes to infinity at the cluster centre. In a recent paper \cite{navarro04} investigated the robustness of the proposed profiles in the centre of clusters and found that both equations are not a good representation in this region. 

Of course, it is very interesting to check whether these proposed profiles fit the overall observed dark matter density profiles in galaxy clusters.

If clusters are roughly spherically symmetric and in hydrostatic equilibrium, which means that no major merger event occurred recently and no shock waves are present, one can use the hydrostatic equation to measure the total mass and the matter distribution of a cluster. The validity of this approach for galaxy clusters has been verified by various studies on hydrodynamic simulations \cite{schindler}, \cite{EMN}.
The hydrostatic equation can be written in the following form:
\begin{equation}
M(r) = - \frac{k T_g(r)r}{\mu m_p G} \left(\frac{\mbox{dln}n}{\mbox{dln}r}
+\frac{\mbox{dln}T_g}{\mbox{dln}r} \right)
\end{equation}
In order to solve the equation it is necessary to know the ICM density and the 
radial temperature distribution.

The ICM density distribution is determined from the X-ray surface brightness distribution either by applying a numerical de-projection\footnote{the gas is optically thin, thus the gas emission is integrated along the line-of-sight. Since the emission goes with the gas density squared, de-projection is not trivial} or by assuming and fitting a model, which can be de-projected analytically.

To obtain the gas temperature profile it is common to perform spectral fits of the ICM emission in concentric annuli around the cluster centre. Before the launch of \xmm and \cha determining the temperature distribution in clusters was very difficult. X-ray observatories, such as {\em ROSAT} or {\em Einstein}, which had the necessary spatial resolution did not have the needed energy range and spectral resolution to  allow an accurate determination of the temperature profile for hot clusters. On the other side, instruments with good spectral capabilities, such as \asca or \sax did not provide the required spatial resolution to allow easily spectro-imaging of clusters. Because of this there was a hot debate whether temperature profiles in clusters decline or not, which has, of course, strong implications on the mass profile.

\xmm and \cha , with their much smaller point spread function make the determination of temperature profiles much easier. Recent results based on \xmm on different clusters such as the Coma cluster \cite{arnaud1}, Abell 1795 \cite{arnaud2}, Abell 1835 \cite{majerowicz}, and Abell 2163 \cite{pratt} are consistent with a flat temperature profile up to half of the virial radius. \cha results of Abell 2163 \cite{markevitch2} are in agreement with the observations by \cite{pratt}. Results on Abell 1835 obtained with \cha by \cite{schmidt} and a study on a sample of galaxy clusters observed with \cha \cite{asf} also suggest a flat temperature profile in outer regions.

A recent paper by \cite{PA1413} on \xmm
data of Abell 1413 suggests a modest temperature decline towards outer regions
and confirm in general the dark matter profiles predicted from above mentioned  numerical simulations. A study on Abell 478 by \cite{pa478} favours clearly the NFW-profile over the Moore et al. profile. This is in agreement with a study on Abell 2029 by \cite{lewis}. These results strengthen generally the idea that dark matter is indeed ``cold''.

\section{Cluster merger in RXJ0256}

\begin{figure}
\psfig{figure=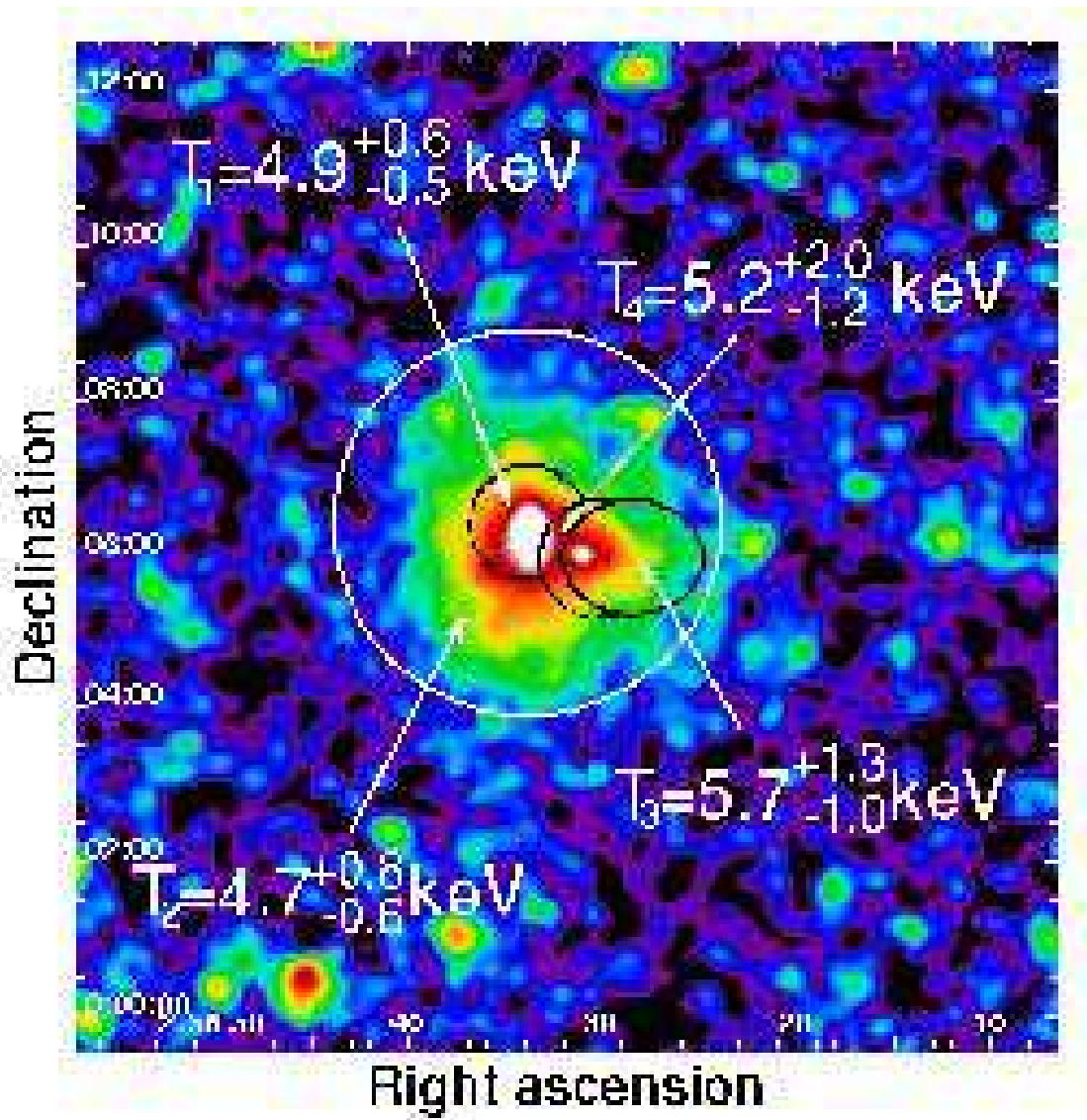,width=15cm}
\caption{X-ray emission image from the 0.3 to 2.0\,keV image of RX~J0256.5+0006 and spectroscopically determined temperature estimates of different cluster regions. The errors are 90\% confidence level.}
\label{fig}
\end{figure}

We studied \xmm data of RX~J0256.5+0006 (z=0.36), a member of the Bright {\sc sharc} --- Serendipitous High-redshift {\sc Rosat} Cluster Survey --- cluster catalog \cite{burke} \cite{collins} \cite{romer} \cite{burke1}. This cluster is one of several {\sc sharc} clusters selected for X-ray follow-up by {\sc Xmm--Newton} (see \cite{arnaud02} \cite{lumb}). Figure \ref{fig} shows the X-ray image of the cluster and the temperature measurements obtained with spectral fitting. The X-ray emitting intra-cluster medium shows a bimodal structure: one main cluster component and a substructure in the west, which very likely falls onto the main cluster centre. The subcluster shows after subtraction of the main cluster component a cometary shape pointing away from the main cluster centre, suggesting that ram pressure stripping is at work. Despite the indication of interaction between the two components we surprisingly do not find any sign of temperature gradients, which is contradictory to predictions from hydro-dynamical simulations of cluster mergers. We develop a simple model to constrain the merger geometry based on an on-axis merger with impact parameter zero. As input we use the projected distance of the subcluster to the main cluster centre observed in X-rays (350~kpc) and estimates of 4 cluster galaxy redshifts obtained with the Kitt Peak telescope. The possible velocity difference between main and subcluster lies between roughly 1800 and 2800~km/s. The subsequent calculated range of possible distances lies between $0.5$~Mpc$ <d<1.4$~Mpc. The relatively large distances of the subcluster to the main cluster centre are confirmed by the fact that we do not see temperature variations along the merger axis (see above). We also examine the effects of ram pressure stripping in the subcluster ICM. We find that the effects we see in X-rays like cometary shape of the subcluster ICM as well as displacement of X-ray maximum with respect to the main galaxy of the subcluster suggest distances to the main cluster centre of about 0.6 to 1~Mpc. This is in very good agreement with our adopted simple merger model and suggests that the impact parameter of the merger is very likely close to zero.

Estimating the mass of the subcluster using a luminosity-mass relationship we find that the ratio of subcluster to main cluster mass is in the order of 20 to 30\%. This ratio is relatively high when compared to other cluster mergers and indicates that we observe a major merger in RX~J0256.5+0006.

Our results obtained for this cluster are described in much more detail in \cite{rxj0256}.

%\bigskip
\bigskip

\vfill
% For Figures insertion uncomment the next section

%\begin{figure}
%\includegraphics{figurename}
%\caption{Your caption here}
%\label{fig01} % optional figure label, must be unique
%\end{figure}

%%%%%%%%%%%%%%%%%%%%%%%%%%%%%%%%%%%%%%%%%%%%%%%%%%%%%%%%
% End of the paper
%
\end{document}